\documentclass[showpacs,pre,aps,twocolumn]{revtex4}  
\usepackage{graphicx,epsfig}
\begin{document}
\title{Turbulence in small-world networks}
\author{M. G. Cosenza}
 \email{mcosenza@ciens.ula.ve}
\author{K. Tucci}
 \email{kay@ula.ve}

 \affiliation{
	$^*$Centro de Astrof\'{\i}sica Te\'orica, Facultad de
             Ciencias, Universidad de Los Andes,   
	Apartado Postal 26 La Hechicera, M\'erida 5251, Venezuela \\
 	$^\dag$SUMA - CeSiMo, Universidad de Los Andes, M\'erida, Venezuela}

\begin{abstract}
The transition to turbulence via spatiotemporal intermittency is 
investigated in the context of coupled maps defined on small-world 
networks. The local dynamics is given by the Chat\'e-Manneville minimal 
map previously used in studies of spatiotemporal intermittency in ordered 
lattices. The critical boundary separating laminar and turbulent regimes 
is calculated on the parameter space of the system, given by the coupling 
strength and the rewiring probability of the network. Windows of 
relaminarization are present in some regions of the parameter space. New 
features arise in small-world networks; for instance, the character of the 
transition to turbulence changes from second order to a first order phase 
transition at some critical value of the rewiring probability. A linear 
relation characterizing the change in the order of the phase transition 
is found. The global quantity used as order parameter for the transition 
also exhibits nontrivial collective behavior for some values of the 
parameters. These models may describe several processes occurring in 
nonuniform media where the degree of disorder can be continuously varied 
through a parameter.
\end{abstract}

\maketitle

\section{Introduction}
A general scenario for the appearance of turbulence in extended systems is
spatiotemporal intermittency, i.e., a sustained regime characterized by the
coexistence of coherent-laminar and disordered-chaotic domains in space and time
\cite{Ciliberto,Kan1}. Some of the most extensive investigations of this phenomenon
have been based on model dynamical systems such as coupled map lattices
\cite{Kan1,Chate1,Chate2,Grass,Stassi}. The idea is that the ingredients of such
models --- a discrete space, discrete time system of interacting elements whose states
vary continuously according to specific functions --- are sufficient to capture much
of the phenomenology observed in complex spatiotemporal processes, in particular some
relevant features of spatiotemporal intermittency and turbulence. In this respect,
coupled map systems can be considered as mathematically simpler and computationally
more efficient models than partial differential equations of hydrodynamics
\cite{Manne}. The transition to turbulence via spatiotemporal
 intermittency has mainly
been investigated in networks of coupled maps where the connectivities between
elements are defined from deterministic rules that provide order to their spatial
structure. There are examples of such studies in Euclidean lattices
\cite{Kan1,Chate1,Chate2,Grass,Stassi}, fractal geometries \cite{CK}, and treelike
arrays \cite{Kay}, as well as on globally coupled systems \cite{Parravano}. These
investigations have allowed the characterization of the transition to turbulence as a
critical phenomenon in a range of Euclidean and fractal dimensions and on several
connection topologies.

Because of their discrete spatial nature, coupled map systems seem especially
appropriate for investigating physical phenomena occurring in heterogeneous or
disordered media. Recently, intensive and interesting research has been performed on
the theory and applications of small-world networks \cite{Watts,Watts2,Mou,Barrat}.
Small-world networks are a class of networks with a high degree of local clustering
and a small characteristic length between any two elements. It has been shown that
small-world networks describe many natural and artificial networks \cite{Newman}. By
varying a parameter, small-world networks can be continuously tuned between ordered,
deterministic lattices and completely random networks. In this article, we consider
coupled maps defined on small-world networks as spatiotemporal dynamical systems. We
study the nature of the transition to turbulence and the properties of spatiotemporal
intermittency on these networks.  We explore the changes induced in those processes as
a result of the variation in the connection topology of the interactions in the
system.

In Sec. II, a general coupled map lattice model for the treatment of small-world
networks is presented. The transition to turbulence in coupled maps on small-world
networks is investigated in Sec. III.  The site map model that we employ is based on
the one introduced earlier by Chat\'{e} and Manneville for regular Euclidean lattices
in one and two dimensions \cite{Chate1}, and which captures the essential features of
the transition to turbulence in extended systems. Section IV contains the observations
of nontrivial collective behavior arising in the system. Conclusions are given in Sec.
V.

\section{Coupled maps on small-world networks}
There are several ways to construct a small-world network. In this article we employ
the small-world network construction algorithm originally proposed by Watts and
Strogatz \cite{Watts}. We start from a ring of $N$ sites, where each site is connected
to its $k$ nearest neighbors, $k$ being an even number. Then each connection is
rewired at random with probability $p$ to any other site of the network, to avoid
self-connections. After the rewiring process, the number of elements coupled to each
site (which we call neighbors of that site) may vary, but the total number of links in
the network is constant and equal to $N k/2$. It is assumed that all links are
bidirectional. Although this algorithm does not guarantee that the resulting graph is
connected, we have used only connected ones for our calculations.

The state of each site on the network can be assigned a continuous variable, which
evolves according to a deterministic rule depending on its own value and the values of
its connecting elements. We define a coupled map lattice dynamical system on a
small-world network as
\begin{equation}
\label{eq_RD}
x_{t+1}(i)= f(x_t(i))+
  \frac{\epsilon}{k(i)} \sum_{j\in S(i)}\left[ f(x_t(j)) - f(x_t(i)) \right];
\end{equation}
where $x_t(i)$ gives the state of the site $i$ $(i= 1,2,\ldots,N)$
at discrete time $t$;
$k(i)$ is the number of neighbors or elements connected to site $i$;
$S(i)$ is the set of
neighbors of site $i$; $\epsilon$ is a parameter measuring the coupling
strength between connected sites, and
$f(x)$ is a nonlinear function describing the local dynamics.

The above coupled map equations can be generalized to include other coupling schemes
or continuous time local dynamics. Different spatiotemporal processes can be studied
on small-world structures by providing appropriate local dynamics and couplings.

\section{Transition to turbulence in small-world networks}
The phenomenon of spatiotemporal intermittency in extended systems consists of a
sustained regime where coherent and chaotic domains coexist and evolve in space and
time. A local map possessing the minimal requirements for observing spatiotemporal
intermittency in coupled map systems is \cite{Chate1}
	
\begin{equation} \label{map}
f(x)=\cases{
	\frac{r}{2}\left(1-\left| 1-2x \right|\right), & if $x \in [0,1]$ \cr
	x, & if $x > 1$, }
\end{equation}
with $r>2$. This map is chaotic for
$f(x)$  in $[0,1]$. However, for $f(x) >1$ the iteration is
locked on a fixed point. The local state can thus be seen as a continuum
of stable ``laminar" fixed points $(x>1)$ adjacent to a chaotic repeller
or ``turbulent" state $(x \in [0,1])$.

In ordered, deterministic networks, the turbulent state can propagate through the
lattice in time for a large enough coupling, producing sustained regimes of
spatiotemporal intermittency. Here, we investigate the phenomenon of transition to
turbulence in small-world networks using the local map $f$ (Eq. (\ref{map})) in the
coupled system described by Eq. (\ref{eq_RD}). The local parameter is fixed at the
value $r=3$ in all the calculations. As in ordered networks, the transition to the
extended turbulent state can be characterized through the average value of the
instantaneous fraction of turbulent sites $F_{t}$, a quantity that acts as the order
parameter for the system \cite{Chate1}. We have calculated  $\langle F \rangle$ as a
function of various parameters of the system from a time average of the instantaneous
turbulent fraction $F_t$, as
\begin{equation}
\label{av}
\langle F \rangle={1 \over T} \sum_{t=1}^T F_t.
\end{equation}
Thus $\langle F \rangle = 0$ describes  a laminar state and $\langle F \rangle \in
(0,1]$ corresponds to a turbulent state of the network. Random initial conditions were
used in all runs with given parameter values of the system Eq.(\ref{eq_RD}). About
$10^4$ iterations were discarded before taking the time average in Eq.(\ref{av}) and
$T$ was typically taken at the value $10^4$. Increasing the averaging time $T$ or the
network size $N$ does not have appreciable effects on the results. It should be
noticed that a minimum number of initially turbulent sites is always required to reach
a sustained state of turbulence.

Figure 1 shows $\langle F \rangle$ as a function of $\epsilon$ for different fixed
values of the probability $p$ in a typical small-world network with 
nearest neighbor number
$k=10$ and size $N=10^4$. The error bars shown on $\langle F \rangle$ correspond to
plus and minus the standard deviations  $\Delta \langle F \rangle$ 
(square root of the variance) of the time
series of $F_t$ at each value of the coupling parameter $\epsilon$. 
We have verified, by doing 25 realizations of the
rewiring process described in Sec. I, that fluctuations on  $\langle F \rangle$
and on $\Delta \langle F \rangle$ 
due to different 
configurations of 
the networks are not significant.  

\begin{figure}[ht]
\centerline{\hbox{
\epsfig{file=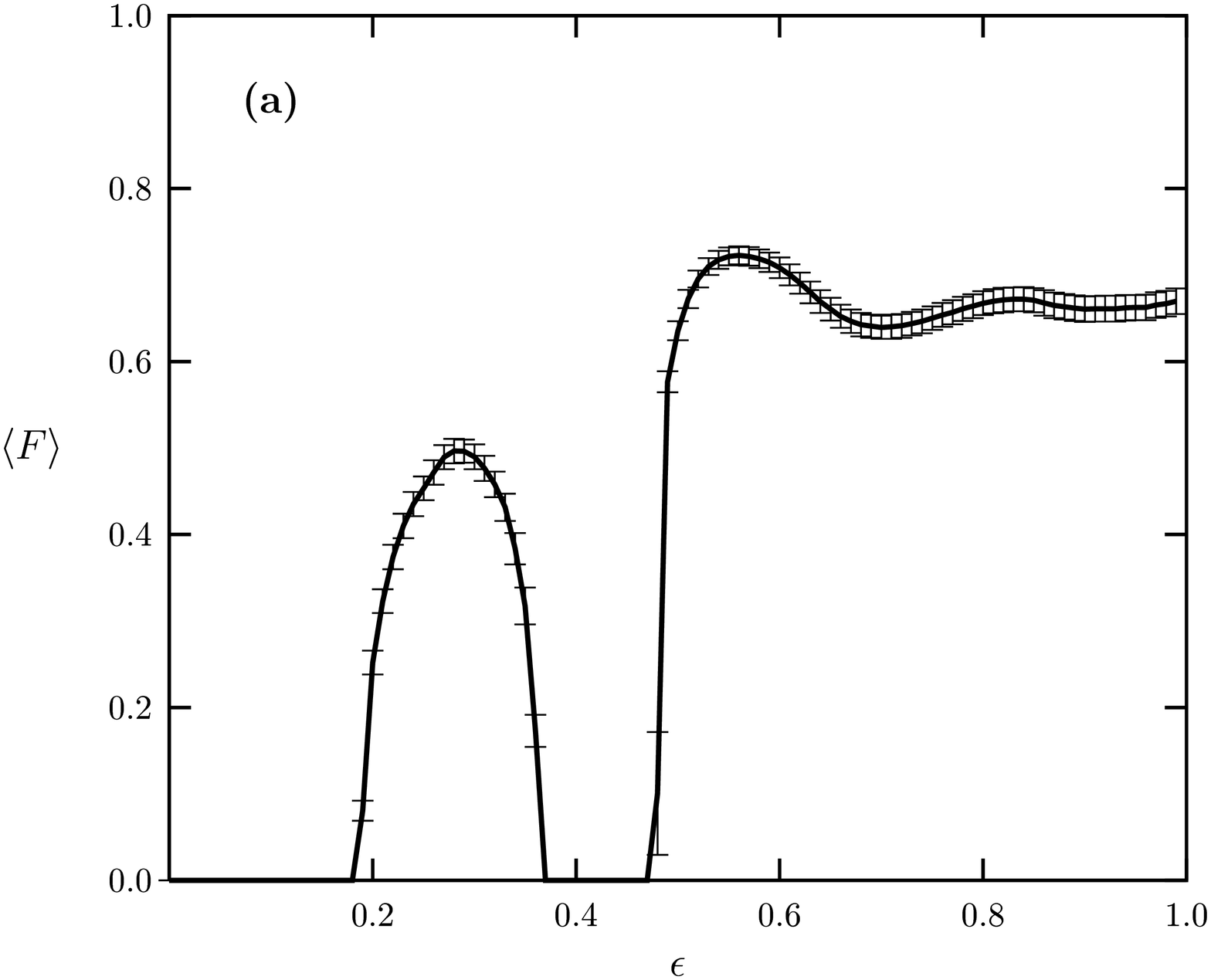,width=.20\textwidth,angle=90,clip=}
\epsfig{file=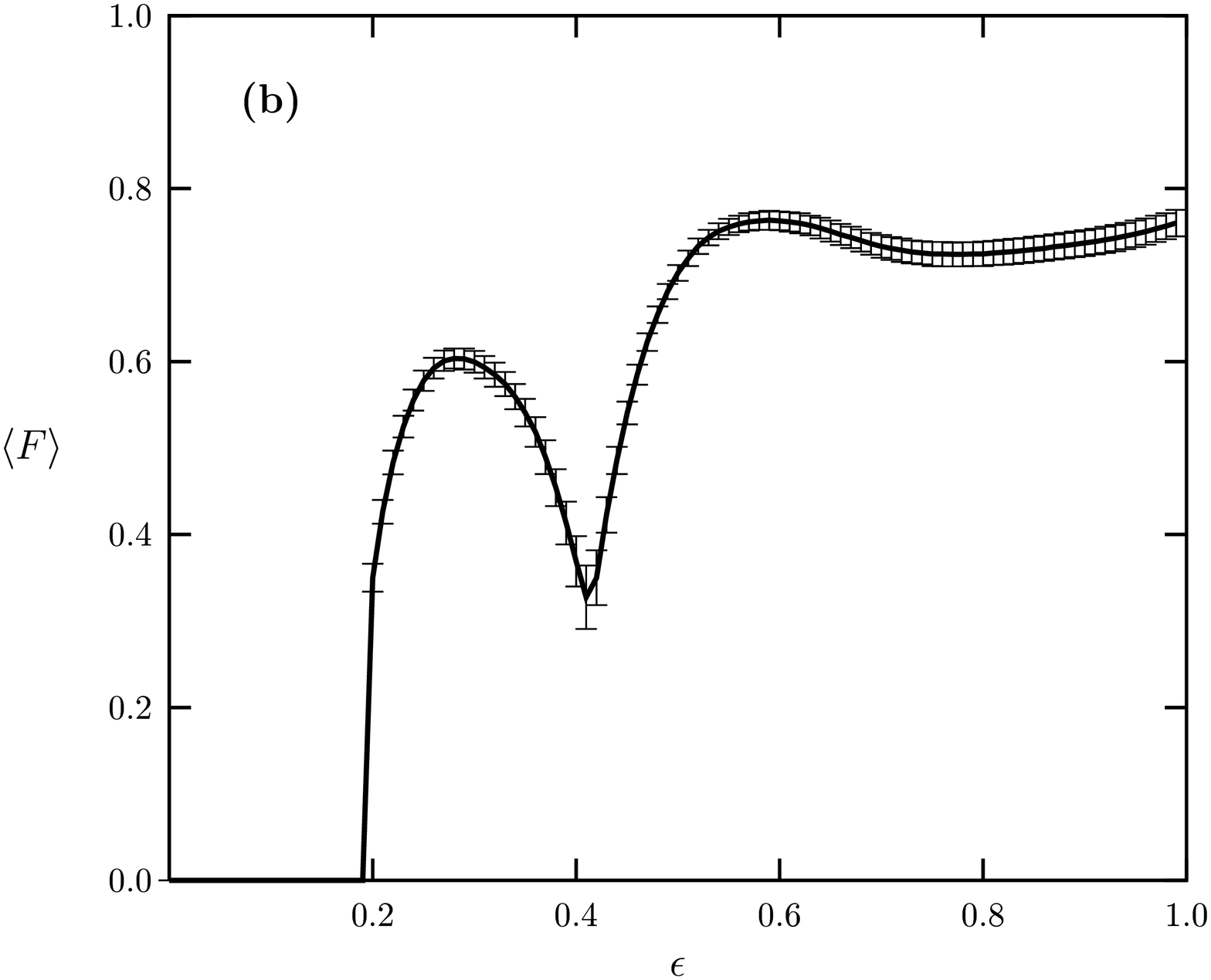,width=.20\textwidth,angle=90,clip=}
}}
\centerline{\hbox{
\epsfig{file=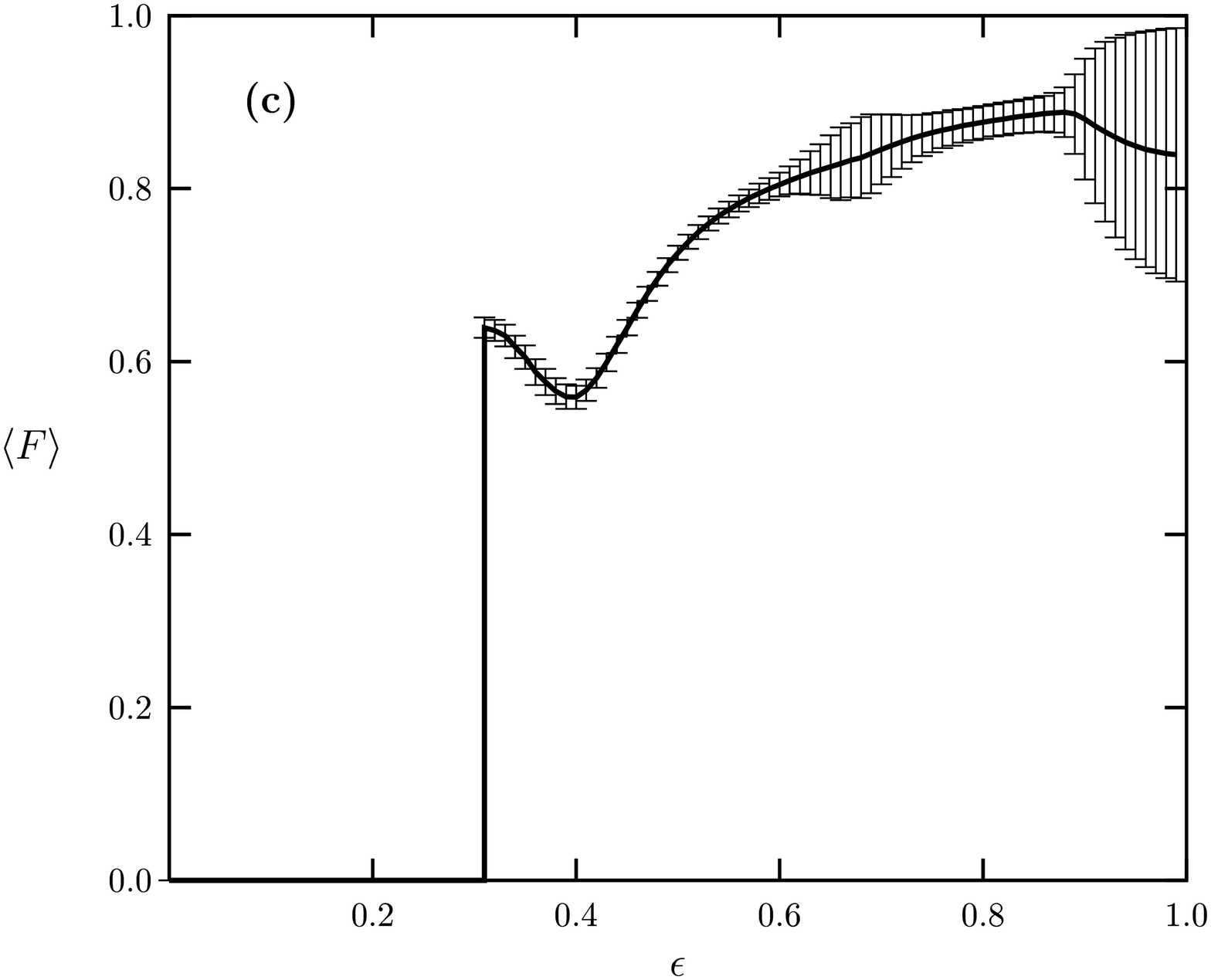,width=.20\textwidth,angle=90,clip=}
\epsfig{file=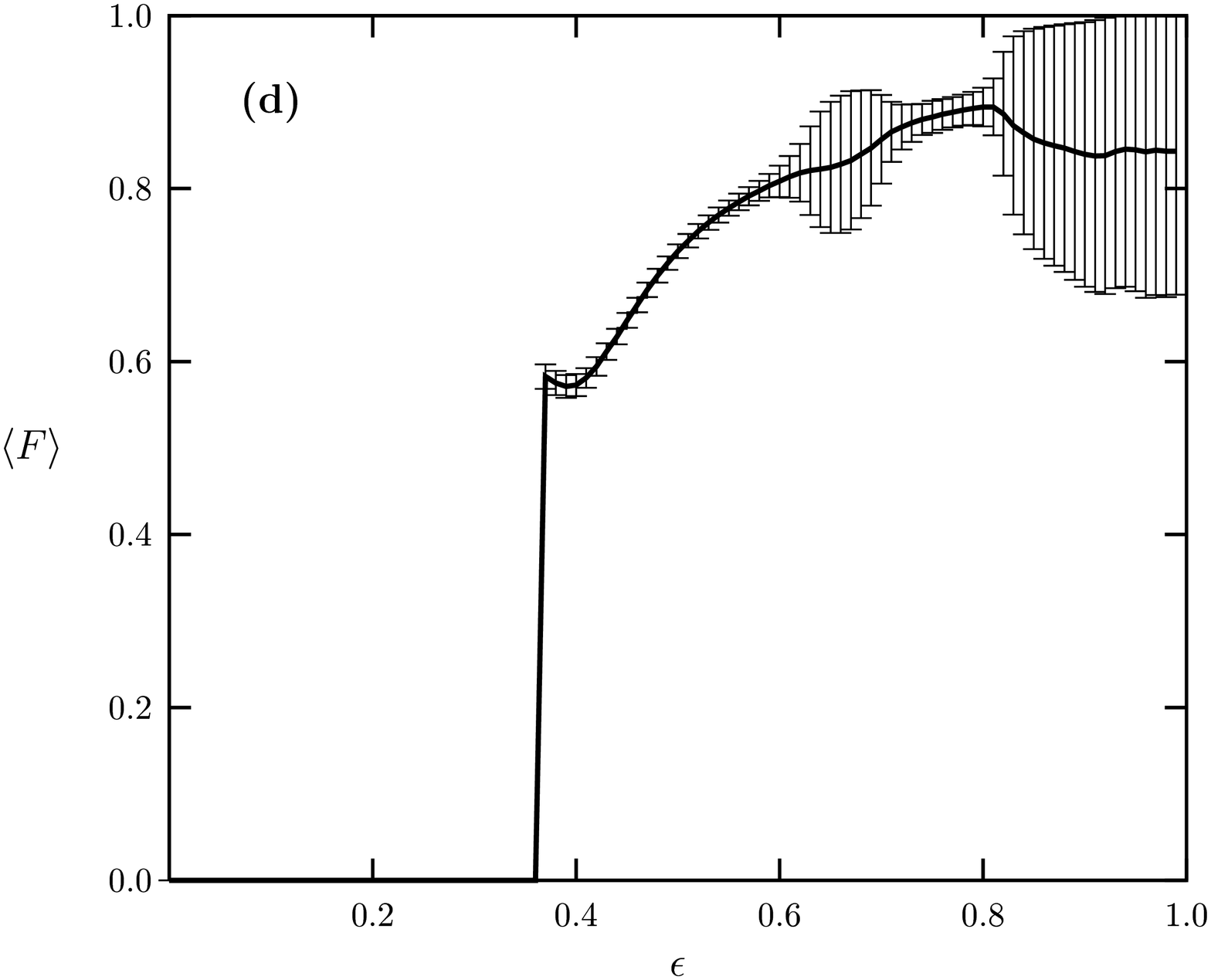,width=.20\textwidth,angle=90,clip=}
}}
\caption{Mean turbulent fraction $ \langle F  \rangle$ as a function of 
the coupling $\epsilon$ for small-world networks with $k=10$ and $N=10^4$. 
The error bars indicate $\pm 1$ standard deviations. (a) $p=0$. (b) $p=0.17$.
(c) $p=p_c=0.55$. (d) $p=0.80$.}
\end{figure}

Figures 1(a)-1(d) show that, for a fixed value
of $p$, there is a critical value of the coupling $\epsilon_c$ at which the transition
from a laminar state to turbulence occurs. For small values of $p$ (Fig. 1(a)), a
regime of relaminarization of the system takes place on an interval of the coupling
parameter, after a window of turbulence. The relaminarization gap disappears with
increasing $p$, leaving a dip in the $\langle F \rangle$ curve, as seen in Fig. 1(b).
For small values of $p$ the transition to turbulence as the coupling is varied takes
place continuously, similarly to a second order phase transition. As $p$ is increased,
the transition becomes progressively steeper until it happens discontinuously, as in a
first order phase transition. There exists a critical value of the probability $p_c
\approx 0.55$ at which the character of the transition to turbulence changes from a
second order phase transition (Figs. 1(a) and 1(b)) to a first order phase transition
(Figs. 1(c) and 1(d)).

In Figs. 2(a) and 2 (b) we show the mean turbulent fraction $\langle F \rangle$ and
its standard deviation $\Delta \langle F \rangle$ plotted as functions of both $p$ and
$\epsilon$. Fig. 2(a) shows that the transition to sustained turbulence occurs on a
critical curve on the parameter plane $(p,\epsilon)$. The variation in the nature of
this transition along this critical curve as $p$ increases can clearly be appreciated
in Fig. 2(a). Typical statistical deviations are seen for small values of $p$ and
$\epsilon$ in Fig. 2(b); however this figure reveals very large fluctuations in the
instantaneous turbulent fraction occurring for larger values of those parameters, and
which can also be observed as the ``bulbs'' in Figs. 1(c) and 1(d). As it shall be
discussed in the next section, this phenomenon is associated to the emergence of
nontrivial collective behavior in the system.

\begin{figure}[ht]
\centerline{\hbox{
\epsfig{file=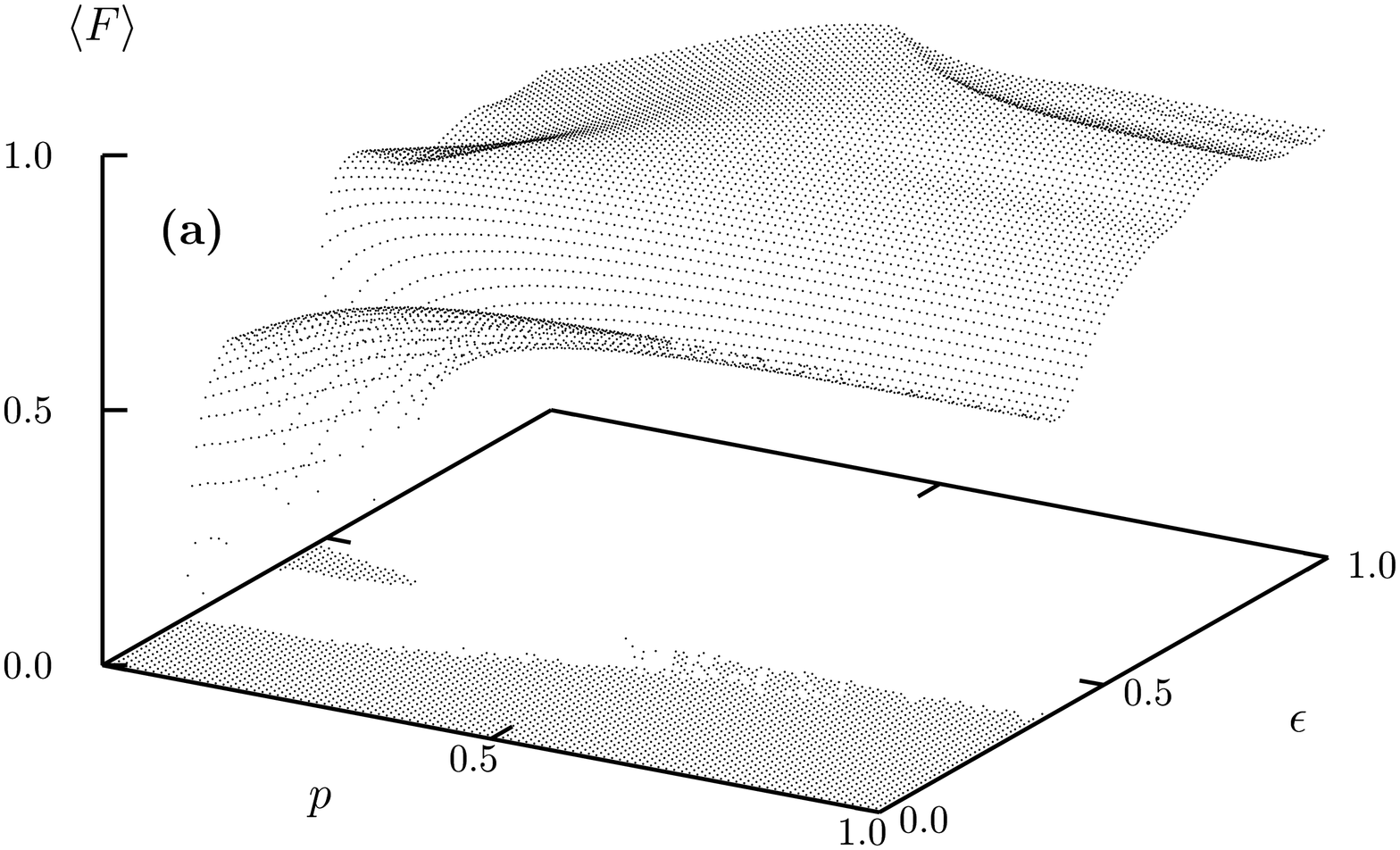,width=.20\textwidth,angle=90,clip=}
\epsfig{file=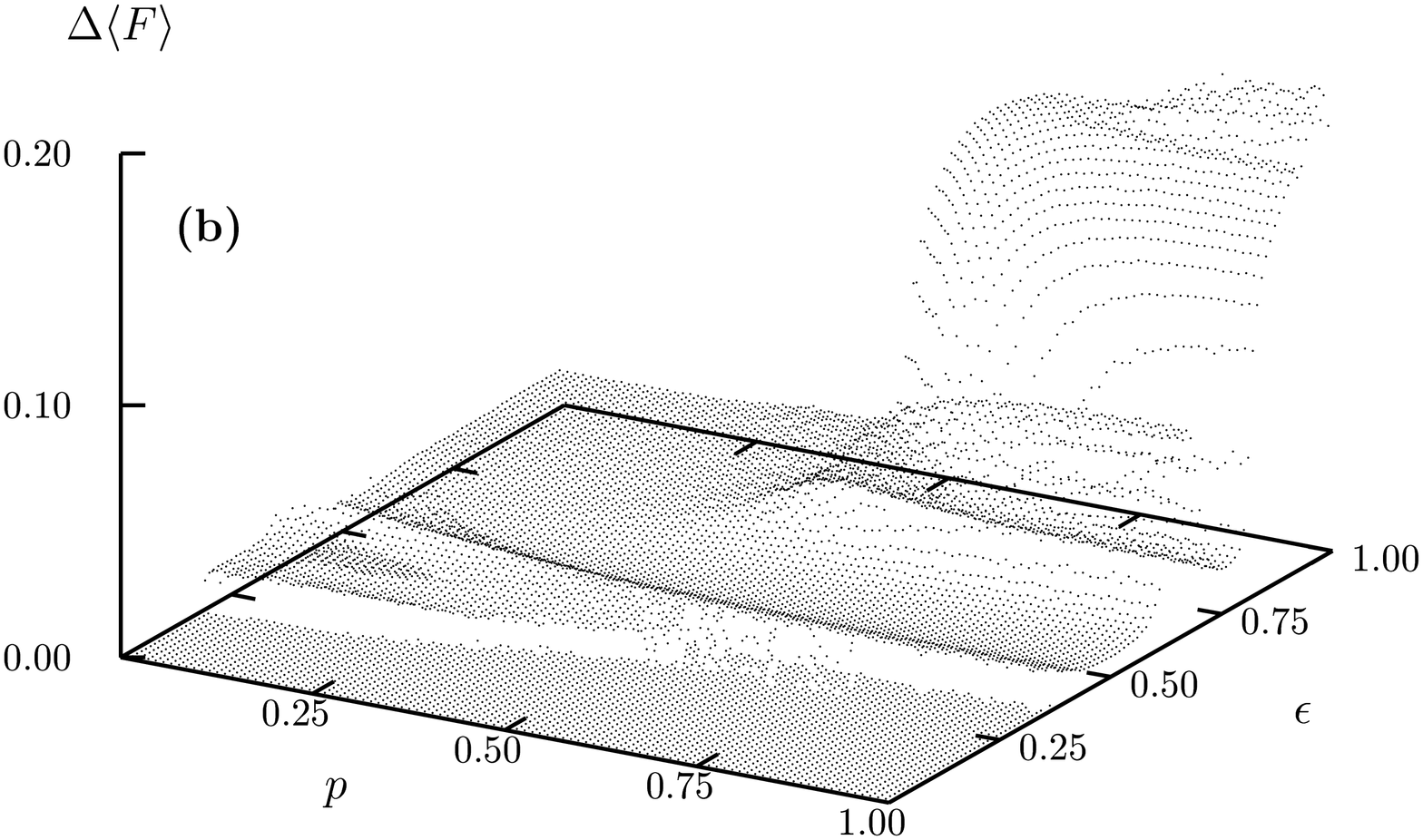,width=.20\textwidth,angle=90,clip=}
}}
\caption{(a) Mean turbulent fraction $\langle F \rangle$
as a function of $p$ and $\epsilon$.
(b) Standard deviation $\Delta \langle F \rangle$ 
as a function of  $p$ and $\epsilon$. Network parameters are
$k=10$, $N=10^4$.}
\end{figure}

We have calculated numerically the critical values of the coupling $\epsilon_c$ for
the onset of turbulence in small-world networks with fixed $k=10$ as a function of
their rewiring probability $p$. Figure 3 shows the resulting critical boundary
$\epsilon_c(p)$ for the transition to turbulence, as well as the phase diagram of the
system, on the parameter plane $(p,\epsilon)$. The critical coupling value
$\epsilon_c$ increases as the disorder in the network, described by the probability
$p$, grows. The critical boundary curve $\epsilon_c(p)$ corresponds to a continuous,
second order phase transition for $p<p_c=0.55$ and to a discontinuous, first order
transition for $p>p_c$. Figure 3 indicates with a dotted line where the first maximum
of the mean turbulent fraction $\langle F \rangle$ occurs on the parameter plane. This
line of first maxima of $\langle F \rangle$ crosses the critical boundary separating
the laminar and the turbulent states of the system at the value $p=p_c$, and the
character of the phase transition changes at this point. The relaminarization gap is
also indicated on the plane $(p,\epsilon)$.

\begin{figure}[ht]
\centerline{\hbox{
\epsfig{file=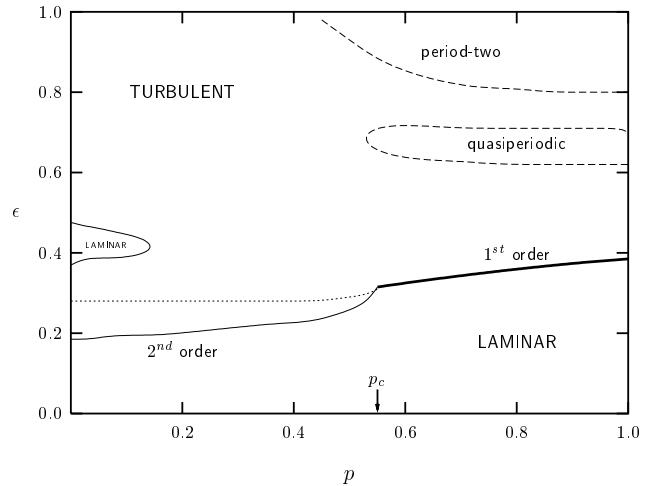,width=.40\textwidth,angle=90,clip=}
}}
\caption{Phase diagram of the system Eq.(\ref{eq_RD}). The critical 
boundary separating laminar and turbulent regimes is shown with a
continuous line. Thin line: second order phase transition; thick line:
first order phase transition. The dotted line indicates the locus of
the first maxima of  $\langle F \rangle$ on the parameter plane.
The regions of nontrivial collective behavior are bounded by dashed
lines.}
\end{figure}

For values of the probability $p<p_c$, where a continuous transition from a laminar
regime to turbulence occurs, the variation of the order parameter $\langle F \rangle$
near the critical curve can be characterized by a critical exponent $\beta$ as
$\langle F \rangle \sim (\epsilon-\epsilon_c)^\beta$. For fixed $p$, the exponent
$\beta$ can be calculated from a log-log plot of $\langle F \rangle$ vs.
$(\epsilon-\epsilon_c)$. The critical exponent $\beta$ varies continuously with the
rewiring probability $p$. In Fig. 4 we show the resulting graphs of $\beta$ as a
function of $p$ for several small-world networks characterized by their neighbor
number $k$. In each case, the dependence of the exponent $\beta$ with $p$ is well
accounted by the linear relation $\beta= h(p_c-p)$, where the slope $h$ varies with
$k$. As $p$ increases, the exponent $\beta$ becomes smaller and the corresponding
phase transition from laminarity to turbulence gets more abrupt. The change in the
character of the transition from second order to first order should occur at the value
$p=p_c$ for which the exponent $\beta$ vanishes. Figure 4 shows the extrapolation of
the straight line corresponding to the small-world network with $k=10$ until its
intersection with the $p$ axis, predicting a critical value $p_c=0.55$. This, in fact,
is the critical value of the probability at which the change in the character of the
transition to turbulence for this network was observed in Figs. 1 (c), 2(a), and 3. We
recall that the measures of the characteristic path length for the family of
small-world networks with $k=10$ drops to small values typically associated to a
random network when the rewiring probability is about $0.55$ \cite{Watts}. Thus the
critical value $p_c=0.55$ found for the emergence of the first order phase transition
should be related to the onset of randomness in the network. In fact, it has recently
been reported that the transition from laminarity to turbulence in randomly coupled
map networks can occur discontinuously \cite{Sandra}.

\begin{figure}[ht]
\centerline{\hbox{
\epsfig{file=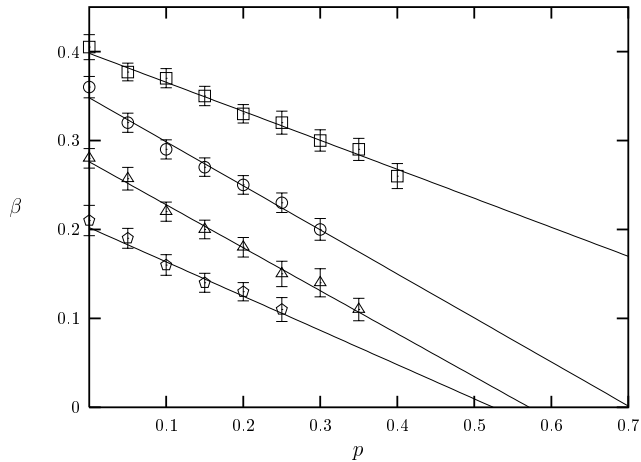,width=.40\textwidth,angle=90,clip=}
}}
\caption{Critical exponent $\beta$ for the second order phase transtion
as a function of the rewiring probability $p$.  $k=4$ (squares);
$k=6$ (circles); $k=10$ (triangles); $k=20$ (pentagons).}  
\end{figure}

By extrapolating the different lines, one can get predictions of the critical
probabilities $p_c$ for different values of $k$. We have numerically verified these
critical values $p_c$ for several small-world networks possessing different neighbor
numbers $k$. Thus the linear relation that arises from Fig. 4 seems to characterize
the transition to turbulence via spatiotemporal intermittency in small-world networks.
Furthermore, Fig. 4 predicts that no critical value $p_c \leq 1$ can be found in the
case $k=4$, and thus the transition to turbulence in a small-world network with
neighbor number $k=4$ should occur continuously (second order) for any $p$, and should
always possess a critical exponent $\beta > 0$. This prediction was also verified
numerically.

Figure 5 shows the predicted critical probability values $p_c$ as a function of $k$.
Note that $p_c$ decreases rapidly with increasing neighbor number $k$. The
characteristic path length becomes smaller with increasing $k$ \cite{Watts} and,
consequently, the connectivity of the small-world network approaches the all-to-all
coupling limit of a globally coupled system, where the transition to turbulence is
always a first order phase transition \cite{Parravano}. In globally coupled systems,
the absence of spatial relations excludes the possibility of supporting small domains
of turbulent maps which would be necessary for a continuous transition to turbulence.
The decrease in the value of $p_c$ observed in Fig. 5 means that the networks need
less disorder for achieving a first order phase transition to turbulence if their
neighbor number $k$ is sufficiently large. This reflects the fact that networks with
large enough $k$ are already closer to global coupling, and therefore they require
less rewiring to behave statistically similar to globally coupled systems.

\begin{figure}[ht]
\centerline{\hbox{
\epsfig{file=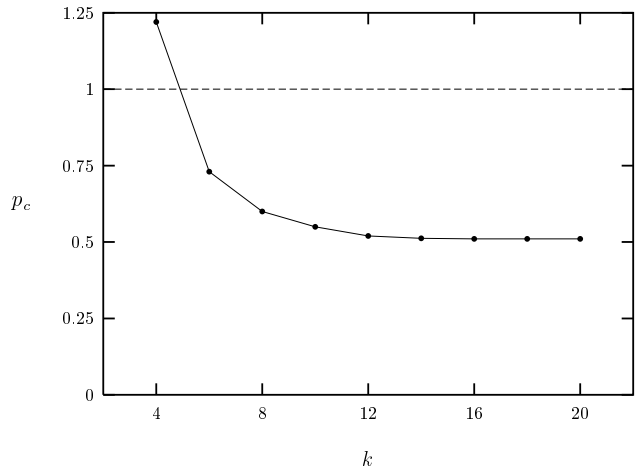,width=.40\textwidth,angle=90,clip=}
}}
\caption{Critical values of the rewiring probability vs. $k$}
\end{figure}

\section{Nontrivial collective behavior}
In contrast to the usually expected statistical behavior, the large amplitudes of the
standard deviations observed on some regions of the parameter plane $(p,\epsilon)$ in
Fig. 2(a) do not diminish with increasing system size or with longer averaging time.
These large fluctuations of a statistical quantity indicate the presence of collective
motions of the system. For example, Fig. 6 displays the return maps of the
instantaneous turbulent fraction $F_t$ for two different values of the coupling
corresponding to the two ``bulbs'' observed in Fig. 1(d). Figure 6(a) shows a
quasiperiodic orbit in the dynamics of $F_t$ in the first bulb, while Fig. 6(b)
reveals a collective period-two motion occurring in the second bulb of Fig. 1(d). The
large fluctuations measured by $\Delta \langle F \rangle$ reflect the amplitude of the
collective oscillations of $F_t$ that emerge in those regions of parameters. This
feature corresponds to a phenomenon of nontrivial collective behavior already observed
in the temporal evolution of the mean field of several chaotic extended systems
\cite{Chate3}. The quantity $F_t$ is a simpler statistical description than the mean
field and, in this case, it already manifests a nontrivial collective behavior of the
system over long times.

\begin{figure}[ht]
\centerline{\hbox{
\epsfig{file=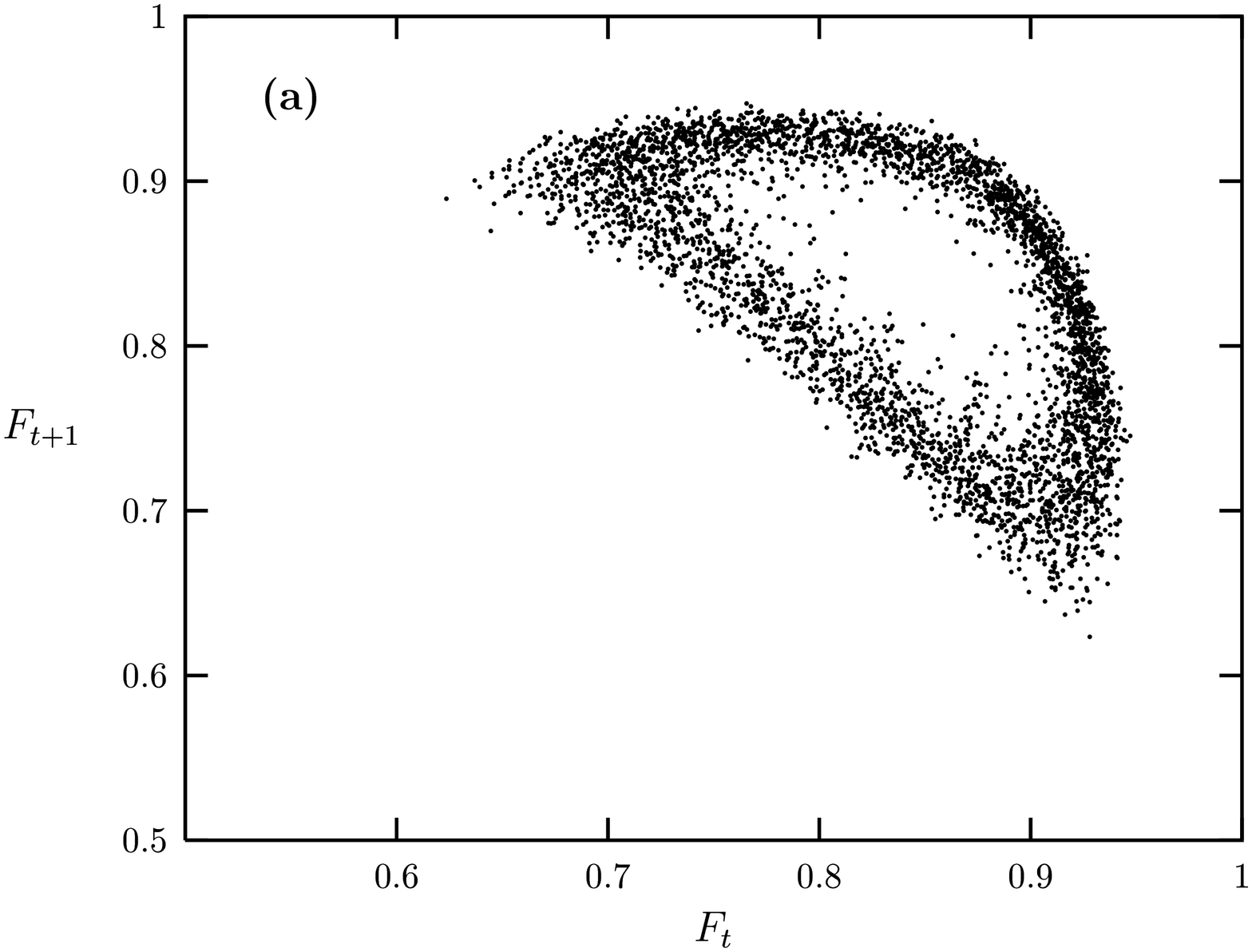,width=.20\textwidth,angle=90,clip=}
\epsfig{file=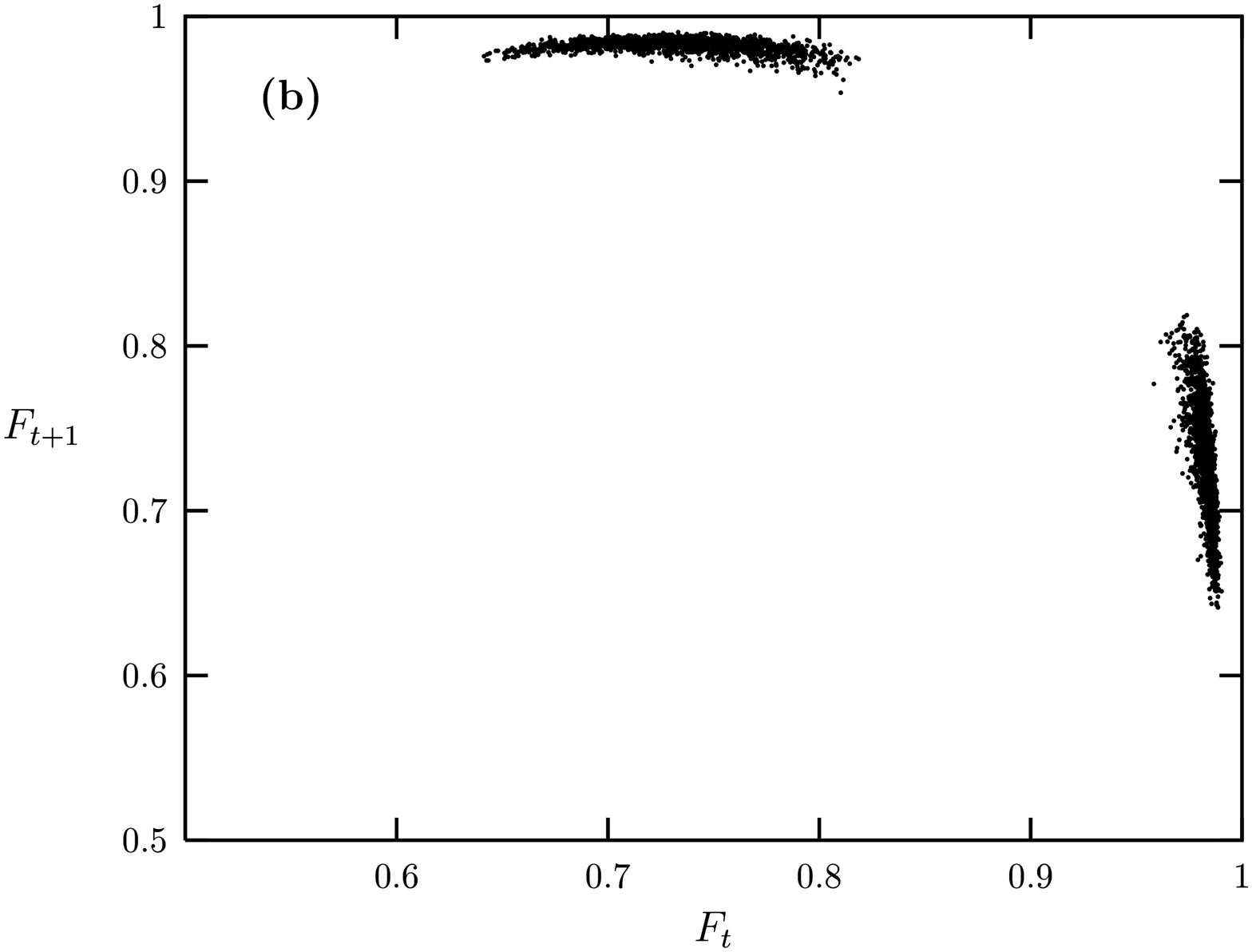,width=.20\textwidth,angle=90,clip=}
}}
\caption{Return maps of the instantaneous turbulent fraction for a small-world
network with $k=10$, $N=10^4$, and $p=0.80$. (a) $\epsilon=0.67$. (b)
$\epsilon=0.85$.}
\end{figure}

The two regions on the parameter plane where states of nontrivial collective behavior
appear are indicated within the turbulent zone in the phase diagram of Fig. 3. Note
that the emergence of nontrivial collective behavior occurs for large enough coupling
and an appreciable degree of disorder, which itself implies a small characteristic
length between any two elements in the network. Thus, as both the amount of short cuts
and the coupling strength between different parts of the network increase, the
information transfer required for the emergence of collective behavior is more likely
to occur.

Nontrivial collective oscillations in the turbulent phase of coupled map systems have
also been observed in high dimensional Euclidean lattices \cite{Chate4}, fractal
lattices \cite{CK}, as well as in globally coupled maps \cite{Parravano}. Our results
for small-world networks show that ordered connections are not essential for the
occurrence of nontrivial collective behavior.

\section{Conclusions}
We have investigated a coupled map model for the transition to turbulence via
spatiotemporal intermittency in small-world networks. Although the local dynamics is simple, we
expect that the essential properties of the transition to turbulence in small-world
structures is captured by this model. Coupled Chat\'e-Manneville maps could be
regarded as a crude description for the dynamics of an excitable medium. The system of
coupled maps on small-world networks can also be used to study different
spatiotemporal dynamical processes on these structures by providing appropriate local
maps or couplings.

By varying the rewiring probability in small-world networks, the behavior of the
transition to turbulence can be studied in the regime between ordered lattices and
completely random networks. The critical boundary separating the laminar and the
turbulent regimes was calculated on the parameter plane $(p,\epsilon)$ of the
system. We have found that the character of the this transition changes progressively
from a second order phase transition to a first order phase transition as the disorder
in the network, measured by the rewiring probability, is increased. The critical value
of the rewiring probability for the onset of the first order phase transition was
predicted from the scaling behavior observed in the critical exponent $\beta$
for small values
of the probability. Additionally, we have been able to calculate the critical values
of the rewiring probability as a function of the number of initial nearest neighbor in
small-world networks.

Discontinuous transition to turbulence and  nontrivial collective behavior in the
turbulent regime are characteristic features of globally coupled maps
\cite{Parravano}. These same collective properties emerge in small-world networks as
their rewiring probability is increased. Because of the ubiquity of small-world
networks in nature and in human-made structures, we may expect to see nontrivial
collective behaviors arising in many systems if the appropriate values of relevant
parameters are reached. We may expect that other typical phenomena of globally coupled
systems, such as the formation of clusters of syncronized elements \cite{Clusters},
could also be observed in small-world networks.

\section*{Acknowledgment}
This work was supported by Consejo de Desarrollo Cient\'{\i}fico, Human\'{\i}stico y
Tecnol\'ogico of Universidad de Los Andes, M\'erida, Venezuela.


\end{document}